\documentclass[usenatbib]{mnras}
\usepackage{graphicx}
\usepackage[export]{adjustbox}
\usepackage{bm}
\usepackage{amsmath}
\usepackage{amssymb}
\usepackage{algorithm}
\usepackage{algpseudocode}
\usepackage{subcaption}
\usepackage{layouts}
\usepackage{hyperref}
\usepackage[capitalise]{cleveref}
\usepackage{fancyvrb}
\usepackage{enumitem}
\usepackage{xcolor}
\usepackage[T1]{fontenc}

\newcommand{\nlive}{n_i}
\newcommand{\Like}{\mathcal{L}}
\newcommand{\DKL}{\mathcal{D}_\mathrm{KL}}
\newcommand{\logLmax}{\log \Like_\mathrm{max}}

\title[\texttt{aeons}]{\texttt{aeons}: approximating the end of nested sampling}
\author[Z. Hu et al.]{Zixiao Hu\thanks{E-mail: zixiao.hu.09@gmail.com},$^{1,2}$  Artem Baryshnikov,$^{1,2}$  Will Handley\thanks{E-mail: wh260@cam.ac.uk}$^{1,2}$
\\
$^{1}$Astrophysics Group, Cavendish Laboratory, J. J. Thomson Avenue, Cambridge, CB3 0HE, UK\\
$^{2}$Kavli Institute for Cosmology, Madingley Road, Cambridge, CB3 0HA, UK
}

\begin{document}
\label{firstpage}
\pagerange{\pageref{firstpage}--\pageref{lastpage}}
\maketitle

\begin{abstract}
    This paper presents analytic results on the anatomy of nested sampling, from which a technique is developed to estimate the run-time of the algorithm that works for any nested sampling implementation. We test these methods on both toy models and true cosmological nested sampling runs. The method gives an order-of-magnitude prediction of the end point at all times, forecasting the true endpoint within standard error around the halfway point.  
\end{abstract}


\section{Introduction}
Nested sampling is a multi-purpose algorithm invented by John Skilling which simultaneously functions as a probabilistic sampler, integrator and optimiser \citep{skilling}. It was immediately adopted for cosmology, and is now used in a wide range of physical sciences including particle physics \citep{Trotta_2008}, materials science \citep{materials} and machine learning \citep{sparse_reconstruction}. The core algorithm is unique in its estimation of volumes by \textit{counting}, coupling together nested monte carlo integrals which makes high-dimensional integration feasible and robust. It also avoids problems that challenge traditional Bayesian samplers, such as posterior multi-modality and phase transitions.
\par
The order of magnitude run-time of an algorithm, that is, whether termination is hours or weeks and months away, is of high importance to the end user. Currently, existing implementations of nested sampling (e.g. \citealt{multinest, polychord, dnest, dynesty, ultranest, nessai,proxnest}) either do not give an indication of remaining run-time, or only provide crude measures of progress that do not directly correspond to the the true endpoint.
\par
\begin{figure}
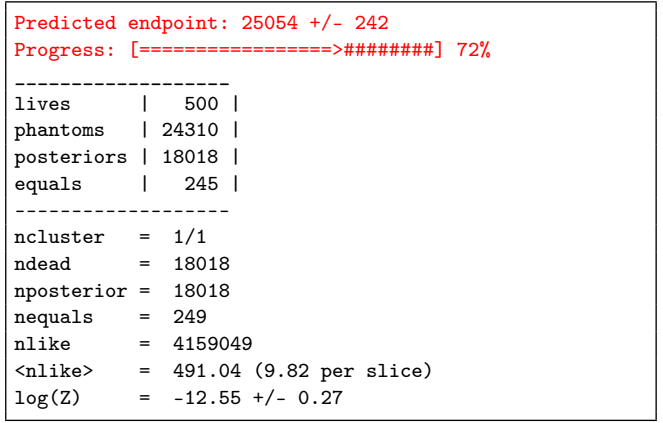

\begin{Verbatim}[frame=single, commandchars=\\\{\}]
\textcolor{red}{Predicted endpoint: 25054 +/- 242}
\textcolor{red}{Progress: [=================>########] 72%}
___________________
lives      |   500 |
phantoms   | 24310 |
posteriors | 18018 |
equals     |   245 |
-------------------
ncluster   =  1/1
ndead      =  18018
nposterior =  18018
nequals    =  249
nlike      =  4159049
<nlike>    =  491.04 (9.82 per slice)
log(Z)     =  -12.55 +/- 0.27
\end{Verbatim}
\caption{Output from \textsc{PolyChord} for a typical nested sampling run. The predicted endpoint, shown in red, is calculated using the method described in this paper.}
\label{fig:polychord_output}
\end{figure}

This paper sets out a principled manner of endpoint estimation for nested sampling at each intermediate stage (as shown in \cref{fig:polychord_output}), the key idea being to use the existing samples to predict the likelihood in the region we have yet to sample from. We begin with an overview of nested sampling in \cref{sec:background}, followed by an examination of the anatomy of a nested sampling run to establish key concepts for endpoint prediction in \cref{sec:anatomy}. \cref{sec:endpoint} then outlines the methodology we use, including discussion and comparisons to previous attempts. Finally, \cref{sec:results} presents the results and discussions for toy and cosmological chains, before we conclude.

\section{Background}\label{sec:background}
We begin with a brief description of the nested sampling algorithm to establish the necessary notation. For a more comprehensive treatment, we recommend the original \citep{skilling} paper, the \citet{sivia} textbook, as well as the excellent technical review by \citet{Buchner_2023} and Nature review by \citet{physical_scientists}. 
\par
\begin{figure*}
\begin{center}
    \includegraphics{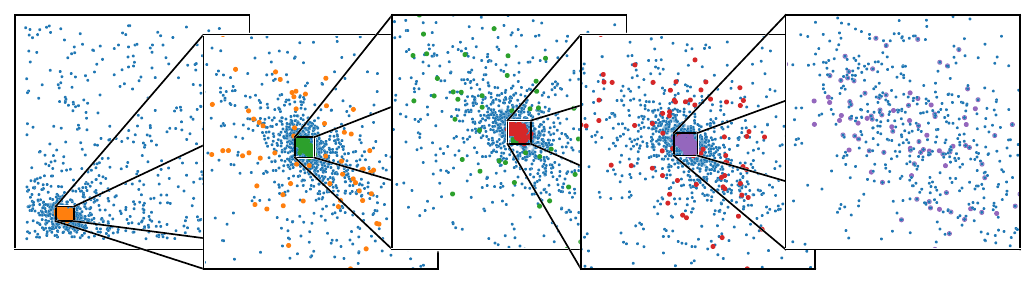}
\end{center}
\caption{The dead points of a nested sampling run, recursively zoomed in. Their density is constant in $\log X$, which is a geometrically constant measure $\propto \mathrm{d} X/X$, and hence scale invariant until the final set of live points is reached. The live points at a given iteration $i$ (larger dots) are uniform across the prior, plotted for comparison.}
\label{fig:dead_measure}
\end{figure*}

For a given likelihood $\mathcal{L}(\theta)$ and prior $\pi(\theta)$, nested sampling simultaneously calculates the Bayesian evidence
\begin{equation}\label{eq:evidence}
	\mathcal{Z} = \int \mathcal{L}\left(\theta\right)\pi(\theta)\ \mathrm{d}\theta
\end{equation}
while producing samples of the posterior distribution
\begin{equation}
	\mathcal{P}(\theta) = \frac{\mathcal{L}(\theta) \pi(\theta)}{\mathcal{Z}}.
\end{equation}
The algorithm operates by maintaining a set of $\nlive$ \textit{live points} sampled from the prior, which can vary in number throughout the run \citep{dynamic_ns}. At each iteration $i$, the point with the lowest likelihood is removed and added to a list of \textit{dead points} (illustrated in \cref{fig:dead_measure}). New points are then (optionally) drawn from the prior, subject to the constraint that they must have a higher likelihood than the latest dead point~$\mathcal{L}_i$. Repeating the procedure leads to the live points compressing around peaks in the likelihood. If there are more births than deaths, the number of live points $n_i$ increases, whilst choosing to not generate new points reduces the number of live points. The exact creation schedule depends on the dynamic nested sampling strategy.
\par
The integral in \eqref{eq:evidence} is then evaluated by transformation to a one-dimensional integral over the \textit{prior volume} $X$
\begin{equation}
	\mathcal{Z} = \int_0^1 \mathcal{L}(X)\ \mathrm{d}X \approx \sum_{i=1} \mathcal{L}_i\ \tfrac{1}{2}(X_{i-1}-X_{i+1}),
\end{equation}
where $X(\mathcal{L})$ is the fraction of the prior with a likelihood greater than $\mathcal{L}$. Whilst the likelihood contour $\mathcal{L}_i$ at each iteration is known, the prior volumes $X_i$ must be statistically estimated as follows: one can define a \textit{shrinkage factor} $t_i$ at each iteration $X_{i} = t_i X_{i-1}$, such that
\begin{equation}\label{eq:X_dist}
	X_i = \prod_{k=1}^i t_k.
\end{equation}
The $t_i$ are the maximum of $\nlive$ points drawn from $[0,1]$, so follow the distribution
\begin{equation}\label{eq:t_dist}
	P(t_i) = \nlive t_i^{\nlive-1},
\end{equation}
\begin{equation}\label{eq:t_moments}
    \langle\log t_i\rangle = -\frac{1}{\nlive}, \quad \mathrm{Var}(\log t_i) = \frac{1}{\nlive^2}.
\end{equation}
The algorithm terminates when an user-specified condition is met; a popular choice is when the evidence in the live points falls below some fraction $\epsilon$ of the accumulated evidence e.g. $10^{-3}$, which is proven to be a valid convergence criterion \citep{evans, Chopin_2010}.
Much of the existing literature treats this remaining evidence separately, for instance by estimating it as the termination $X$ multiplied by the average likelihood amongst the remaining live points. It is, however, quantitatively equivalent but qualitatively neater to consider termination as killing the remaining live points off one-by-one, incrementing the evidence exactly as during the run with decreasing $\nlive$ \citep{dynesty}.
\par
Uncertainties in the evidence are dominated by the spread in the prior volume distribution, and the simplest way to estimate them is by Monte Carlo sampling over sets of $\bm{t}$. As Skilling and others \citep{Chopin_2010, Keeton_2011} have shown, for any given problem the uncertainty in $\log \mathcal{Z}$ is proportional to $1/\sqrt{n_\mathrm{live}}$, so $n_\mathrm{live}$ sets the resolution of the algorithm.   

\section{The anatomy of a nested sampling run}\label{sec:anatomy}
The following sections act as an inventory of the information available to us at an intermediate iteration $i$, which we shall use to make endpoint predictions in \cref{sec:endpoint}. We present an anatomy of the progression of a nested sampling run in terms of the prior volume compression (\cref{sec:prior_volume}), the log-likelihood increase (\cref{sec:logL}), the inferred temperature (\cref{sec:temperature}), and the dimensionality of the samples (\cref{sec:dimensionality}).

\subsection{Prior volume}\label{sec:prior_volume}
The key feature of nested sampling is that the sampling is controlled by prior volume compression. The task is to find the posterior typically lying in a tiny fraction of the prior volume, a total compression which is quantified by the average information gain, or \textit{Kullback-Leibler divergence}:
\begin{equation}\label{eq:DKL}
   \DKL = \int \mathcal{P}(\theta) \log \frac{\mathcal{P}(\theta)}{\pi(\theta)}\ \mathrm{d}\theta. 
\end{equation}
The bulk of the posterior lies within a prior volume ${X = e^{-\DKL}}$, which is the target compression. From \cref{eq:t_moments} one gets there by iteratively taking steps of size ${\Delta \log X_i = -1/n_i}$, so that when we add up the contribution of each step in \eqref{eq:t_moments} we get
\begin{equation}
    \langle\log X_i\rangle = -\sum_{k=1}^i \frac{1}{n_k}, \quad \mathrm{Var}(\log X_i) = \sum_{k=1}^i \frac{1}{n_k^2}.
\end{equation}
A steady step size in $\log X$ corresponds to a geometrically constant measure for the dead points, which is exactly needed to overcome the curse of dimensionality.
\par
The same is not true for the live points, which are uniformly distributed in prior volume, and by the comments in the penultimate paragraph of \cref{sec:background} have ${n_{i+k} = n_i-k}$. As a result, the maximum live point is found at 
\begin{equation}\label{eq:Xmin}
    \langle\log X_\mathrm{min}^{\mathrm{live}}\rangle = \langle\log X_i\rangle - \sum_{k=1}^{\nlive} \frac{1}{n_{i+k}} \approx -\frac{i}{\nlive} - \log \nlive - \gamma,
\end{equation}
with variance
\begin{equation}
    \mathrm{Var}(\log X_\mathrm{min}^{\mathrm{live}}) = \mathrm{Var}(\log X_i) + \sum_{k=1}^{\nlive} \frac{1}{n_{i+k}^2} \approx \frac{i}{\nlive^2} + \frac{\pi^2}{6}, 
\end{equation}
where the large $\nlive$ limit is taken for the approximation to the harmonic series, $\gamma$ being the Euler-Mascheroni constant.
\par
The live points therefore only get us a factor of $\log \nlive$ closer in volume to the posterior bulk. In other words, it is not until we are around $\log \nlive$ away from $\log X = -\DKL$ that the samples begin populating the posterior typical set. One can see from \eqref{eq:DKL} that the divergence increases linearly with dimension, so for large dimensionalities and typical live point numbers $\lesssim 1000$, this does not happen until near the end of the run.
\par
\begin{figure*}
\begin{center}
    \includegraphics{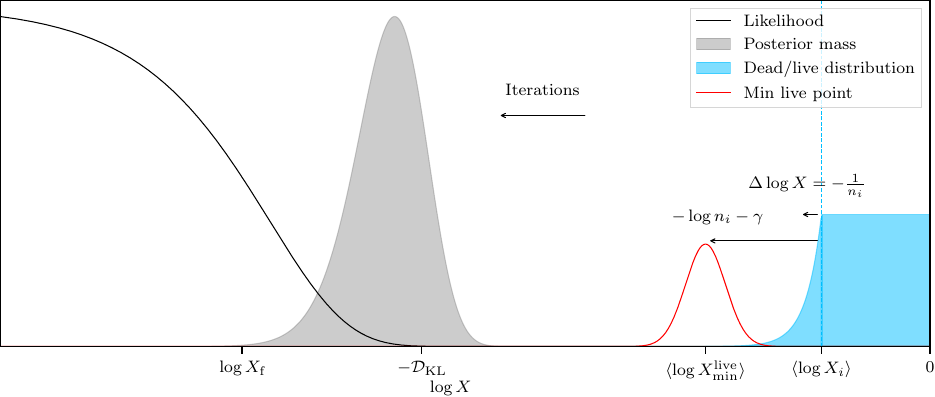}
\end{center}
\caption{The distribution of the posterior mass in terms of $\log X$, the live points over the constrained prior and the smallest live point prior volume $\log X_\mathrm{min}^{\mathrm{live}}$ at an intermediate iteration $i$. For large values of $\DKL$ i.e. informative posteriors and/or large dimensionalities, the maximum live point is very far from the posterior bulk until the very end of the run. Note that the x-axis in this plot is $\log X$, so that the run proceeds from right to left to emphasise that the enclosed prior volume iteratively gets smaller. Plots of the sort from here onward will be in terms of $-\log X$, where the run will more naturally proceed from left to right. }
\label{fig:logX_distribution}
\end{figure*}

The result is consistent with that in \citet{statmech}, which states that a spike at a volume smaller than $X_i/n_i$ will go undetected. Intuitively, it is because for a sharply peaked likelihood the live points are too diffuse to land there with any significant probability for most of the run. These results are summarised in \cref{fig:logX_distribution}

\subsection{Log-likelihood}\label{sec:logL}
We now consider the distribution of the samples in log-likelihood. To get an insight into the analytics, we will examine the representative case of the $d$-dimensional multivariate Gaussian with maximum point $\logLmax$ and length scale $\sigma$:
\begin{equation}\label{eq:gaussian_logL}
    \log \Like = \logLmax - X^{2/d}/2\sigma^2.
\end{equation}
Providing that $\sigma^d \ll 1$, the posterior can be approximated as 
\begin{equation}
    \mathcal{P}(X) = \tfrac{1}{(2\pi\sigma^2)^{d/2}} e^{-X^{2/d}/2\sigma^2}, 
    \label{eqn:PX}
\end{equation}
which in terms of $\log\mathcal{L}$ becomes
\begin{equation}
    \mathcal{P}(\log\mathcal{L}) = \frac{1}{\Gamma\left(\frac{d}{2}\right)}e^{\log\mathcal{L}-\log\mathcal{L}_\mathrm{max}} (\log\mathcal{L}_\mathrm{max}-\log\mathcal{L})^{\frac{d}{2}-1}
\end{equation}
i.e. $2(\log\mathcal{L}_\mathrm{max}-\log\mathcal{L}) \sim \chi^2_{d}$, with mean and variance
\begin{equation}
    \langle\log\mathcal{L}\rangle_\mathcal{P} = \log\mathcal{L}_\mathrm{max} - \frac{d}{2},  \quad \mathrm{Var}(\log\mathcal{L})_\mathcal{P} = \frac{d}{2}.
\end{equation}
Given the prior is uniform $X\sim U(0,1)$, the KL divergence in the $\sigma^d \ll 1$ limit is:
\begin{equation}
\mathcal{D}_\mathrm{KL} = -\frac{d}{2}\log 2 \Gamma(1+\tfrac{d}{2})^{2/d} \sigma^2.
\end{equation}

One can also derive the individual distributions for the live and dead points in log-likelihood. Note that this is merely the distribution of the \textit{values} of the points, without their nested sampling weights. The dead distribution is
\begin{equation}
    P(\log\Like) = \frac{\mathrm{d} \log X}{\mathrm{d} \log\Like} P(\log X) \propto \frac{d\:n_i}{2(\logLmax - \log\Like)},
\end{equation}
since the log-densities of the dead point prior volumes are proportional to the live point number at any given iteration. Meanwhile, the live points are uniformly distributed, and so have the distribution 
\begin{multline}
	P(\log\mathcal{L}) = \frac{d}{2}\frac{(\log\mathcal{L}_\mathrm{max}-\log\mathcal{L})^{\frac{d}{2}-1}}{(\log\mathcal{L}_\mathrm{max}-\log\mathcal{L}_i)^{\frac{d}{2}}} \\
    [\log\mathcal{L}_i < \log\mathcal{L} <\log\mathcal{L}_\mathrm{max}].
    \label{eq:PL}
\end{multline}
\cref{fig:logL_distribution} shows the distinction between the distributions of the live and dead point values, as well as the posterior. 
\par
How much further do the live points penetrate in log-likelihood? We seek the distribution for the maximum likelihood of the live points, $\log\Like_{(n_i)}^{\mathrm{live}}$, where we have used the notation of order statistics to denote $x_{(k)}$ as the maximum of $k$ points.  For convenience, we will normalise the likelihoods as
\begin{equation}
    y = \frac{\log\mathcal{L}-\log\mathcal{L}_i}{\log\mathcal{L}_\mathrm{max}-\log\mathcal{L}_i}
    \label{eq:normalised_likelihood}
\end{equation}
so that $y=0$ corresponds to the latest dead point at $\log\Like_i$ and $y=1$ to the maximum. The live point distribution simplifies to
\begin{equation}
    P(y) = \frac{d}{2}(1-y)^{\frac{d}{2}-1} \quad [0<y<1].
    \label{eq:Py}
\end{equation}
Using the result that the maximum of $n_i$ variables with cumulative distribution $F(y)$ follows $\frac{d}{dy}( 1- (1-F(y))^{n_i})$, we obtain
\begin{multline}
    P(y_{(n_i)}^\mathrm{live}) = \frac{n_i d}{2}(1-y_{(n_i)}^\mathrm{live})^{\frac{d}{2}-1}\left(1-(1-y_{(n_i)}^\mathrm{live})^{\frac{d}{2}}\right)^{n_i-1}\\ 
    [0<y_{(n_i)}^\mathrm{live}<1],
    \label{eq:Pyhat}
\end{multline}
which may be roughly summarised as
\begin{multline}
    y_{(n_i)}^\mathrm{live} \sim 1-\frac{\Gamma(1+\frac{2}{d})\Gamma(1+n_i)}{\Gamma(1+\frac{2}{d}+n_i)} \\
     \pm \left( \frac{\Gamma(1+n_i)\Gamma(1+\frac{4}{d})}{\Gamma(1+\frac{4}{d}+n_i)} - \frac{\Gamma(1+\frac{2}{d})^2 \Gamma(1+n_i)^2}{\Gamma(1+\frac{2}{d}+n_i)^2}\right)^{\frac{1}{2}},
    \label{eq:ymax}
\end{multline}
or in the large $d$, $n_i$ limit
\begin{align}
    \lim_{d\to\infty} y_{(n_i)}^\mathrm{live} &\sim \frac{2H_{n_i}}{d} \pm \left(\frac{2(\pi^2 - 6\Psi^{(1)}(1+n_i))}{3d^2}\right)^{\frac{1}{2}},
    \label{eq:ymaxd}\\
    \lim_{d,n_i\to\infty} y_{(n_i)}^\mathrm{live} &\sim \frac{2\log n_i}{d} \pm \sqrt{\frac{2}{3}}\frac{\pi}{d},
    \label{eq:ymaxdn}
\end{align}
where $\psi^{(1)}$ is the trigamma function and $H_{n_i}$ is the $n_i$th harmonic number. This is a very small fraction in high dimensions, showing that until the end the live points are far from the posterior bulk. 
\par
Alternatively, the same result arrives intuitively from the fact that at each step, $y$ increases by  
\begin{equation}\label{eq:dy}
    \lim_{\nlive \to \infty}\Delta y \approx \frac{\mathrm{d} y}{\mathrm{d}\log X} \Delta \log X = \frac{2}{d \nlive}, 
\end{equation}
so that by again summing the harmonic series, we get
\begin{equation}
    y_{(\nlive)}^{\mathrm{live}} = \frac{2\log \nlive}{d}.
\end{equation}
\cref{eq:dy} also implies that the normalised distance between the highest and second highest live point is roughly 
\begin{equation}
    y_{(\nlive)}^{\mathrm{live}} - y_{(\nlive - 1)}^{\mathrm{live}} \approx \frac{2}{d}.
\end{equation}
Before reaching the posterior bulk, $\logLmax - \log\Like_i > d/2$, so we must have
\begin{equation}
    \log \Like^{\mathrm{live}}_{(\nlive)} - \log \Like^\mathrm{live}_{(\nlive - 1)} > 1.
\end{equation}
In other words, the highest likelihood point is always at least an order of magnitude greater than the second highest. It is therefore typically the case that nearly all of the posterior mass is concentrated in a single point, the maximum live point, until the very end of the run when the prior volumes have shrunk enough to compensate.
\subsubsection*{Aside: nested sampling as a maximiser}
Previous literature \citep{Akrami_2010, Feroz_2011} has explored the potential for nested sampling to be used as a global maximiser, given its ability to handle multi-modalities. In particular, the latter authors emphasised that posterior samplers such as nested sampling find the bulk of the \textit{mass}, not the maximum of the distribution, but that this can be remedied by tightening the termination criterion. We now use the machinery we have developed to put this statement on a more quantitative footing. 
\par
Let us take the current iteration to be the termination point with likelihood $\log\Like_\mathrm{f}$ and prior volume $X_\mathrm{f}$, so that
\begin{equation}
	\epsilon = \frac{\int_0^{X_\mathrm{f}} \mathcal{L}\ \mathrm{d}X}{\int_0^\infty \mathcal{L}\ \mathrm{d}X}.
\end{equation}
Note that we have assumed that prior effects are negligible (so $1\approx \infty$), and that $\epsilon \ll 1$ so that the denominator is approximately the accumulated evidence. Computing this for \eqref{eq:gaussian_logL}, we find the answer in terms of lower incomplete gamma functions
\begin{equation}
\epsilon = 1- \frac{\Gamma_{d/2}\left(X_\mathrm{f}^{2/d}/2\sigma^2\right)}{\Gamma(d/2)}.
\end{equation}
Taking the $X_\mathrm{f}\ll (\sqrt{2}\sigma)^d$ limit (almost certainly valid at termination) we find
\begin{equation}
    \lim_{X_\mathrm{f}\ll (\sqrt{2}\sigma)^d} \epsilon \approx \frac{X_\mathrm{f}}{(\sqrt{2}\sigma)^d \ \Gamma\left(1+\frac{d}{2}\right)} = \frac{(\log\mathcal{L}_\mathrm{max}-\log\mathcal{L}_\mathrm{f})^{\frac{d}{2}}}{\Gamma\left(1+\frac{d}{2}\right)}.
\end{equation}
We thus have an expression relating $\mathcal{L}_\mathrm{f}$ at termination to the termination fraction $\epsilon$. This becomes yet more pleasing in the large $d$ limit, since $\epsilon^{2/d}\to 1$, we find via a Stirling approximation:
\begin{equation}
    \lim_{d\to\infty} \log\mathcal{L}_\mathrm{f} \approx \log\mathcal{L}_\mathrm{max} - \frac{d}{2e}.
\end{equation}
In the event that we retain $\epsilon$, we replace $\frac{d}{2e}\to \frac{d}{2e}\epsilon^{2/d}$, allowing one to battle the $\frac{d}{2e}$ term exponentially as dimensions increase.
\par
Putting this together, taking $\mathcal{L}_i$ in \eqref{eq:normalised_likelihood} to be $\mathcal{L}_\mathrm{f}$ and combining this with \eqref{eq:ymaxdn}, we find
\begin{equation}
    \boxed{
        \log{\mathcal{L}}_\mathrm{max}^\mathrm{live} \approx \log\mathcal{L}_\mathrm{max} - \frac{d}{2e} + \frac{\log n_i}{e} \pm \frac{\pi}{\sqrt{6}e}
    },
\end{equation}
showing that in general nested sampling will finish at a contour $d/2e$ away from the maximum log-likelihood. The final set of $n_i$ live points gets you $\log n_i/e$ closer, with a chance of getting $\sim\pi/\sqrt{6}e=0.472$ closer still by statistical fluctuation. 
\par
Making the traditional termination criterion stricter therefore has limited returns in high-dimensions, if it is ultimately still based on the remaining evidence. However, nested sampling still shrinks around the maximum exponentially, so provided a good alternative termination criterion is chosen, it will get there in reasonable time. 

To quantify this statement, consider the number of iterations $\Delta i$ required to get from the posterior bulk to the true maximum, which we will now calculate. At the posterior, we have roughly speaking
\begin{equation}
    (\log X, \log\Like) = (-\DKL, \logLmax - d/2).
\end{equation}
The prior volume $\log X_\delta$ at which the likelihood is within $\delta$ of the maximum can then be found by inverting \cref{eq:gaussian_logL}:
\begin{equation}
    \log X_\delta = -\DKL - \frac{d}{2} \left(\log \frac{d}{2} - \log \delta\right),
\end{equation}
which corresponds to an additional
\begin{equation}
    \Delta i = \frac{nd}{2} \left(\log \frac{d}{2} - \log \delta\right)
\end{equation}
iterations after the bulk is reached, where $n$ is the harmonic mean of the number of live points. One would therefore expect a general distribution to take $\mathcal{O}(\tfrac{n d }{2}\log \tfrac{d}{2})$ iterations to get from the usual nested sampling stopping point to within an $e$-fold of the maximum. A rule-of-thumb termination criterion could therefore be to run for at least $\tfrac{nd}{2}\log\tfrac{d}{2}$ iterations after the posterior is reached.
\par
\begin{figure*}
\begin{center}
    \includegraphics{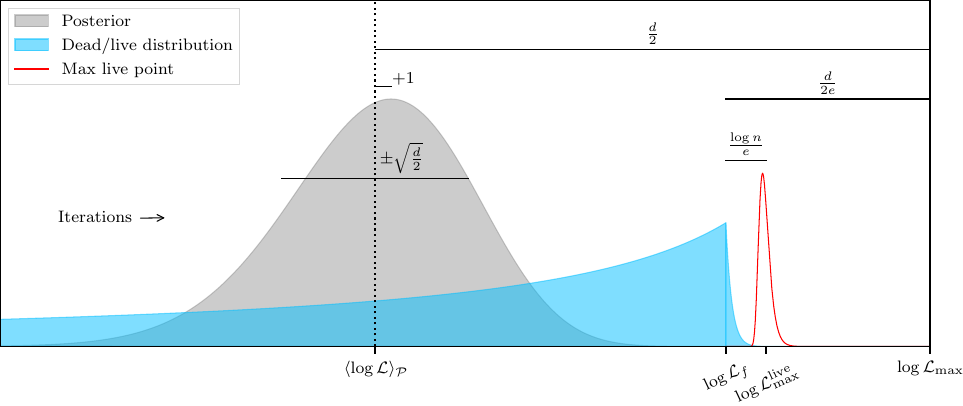}
\end{center}
\caption{Distribution of samples as a function of $\log\mathcal{L}$, showing the posterior $\mathcal{P}(\log\mathcal{L})$, the distribution of the live points $\mathcal{\pi}(\log\mathcal{L} \mid \mathcal{L}>\mathcal{L}_i)$, and the distribution of the maximum likelihood live point $P(\log\mathcal{L}_\mathrm{max}^\mathrm{live})$. The distances are shown between these locations at the end of the run, the key takeaway being that in high dimensions the highest log-likelihood point of a nested sampling run is nowhere near the maximum in high dimensions.}
\label{fig:logL_distribution}
\end{figure*}

A summary of the distances between the notable points at the end of a run is shown in \cref{fig:logL_distribution}.

\subsection{Temperature}\label{sec:temperature}
\subsubsection*{Motivations}
As shown in the previous section, midway through the run nearly all of the posterior mass is concentrated at a single point. However, this does not capture the \textit{structure} of the posterior that has been explored and all of the information it provides. 
\par
\begin{figure}
\begin{center}
    \includegraphics{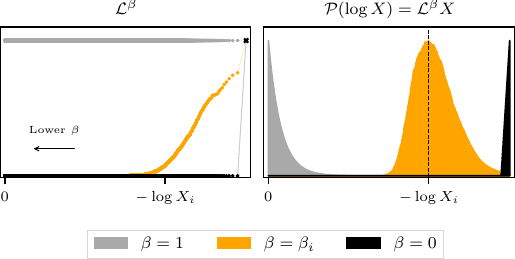}
\end{center}
\caption{The likelihood and posterior as a function of $\log X$ in the middle of a nested sampling run. Almost all of the posterior mass is concentrated at a single point with the highest compression, because it is orders of magnitude higher in likelihood. Reducing $\beta$ re-weights to shift the posterior mass; taking $\beta \to 0$ goes too far and gives back the prior, but there is some intermediate $\beta = \beta_i$ which is significant because it centres the posterior at the current contour.}
\label{fig:last_live_point}
\end{figure}

We have the potential to fix this because nested sampling is invariant to monotonic transformations, so we can transform the likelihood as $\Like \to \Like^{\beta}$ without loss of information by trivially re-weighting the samples. Increasing $\beta$ worsens the situation, while $\beta \to 0$ simply gives back the prior. There is, on the other hand, a significant intermediate range which makes the samples look like a posterior centred at the present contour, which will allow us to recover the structure of the samples. A schematic of the procedure is shown in \cref{fig:last_live_point}.
\par
\par
At this point it is relevant to note the correspondence between Bayesian inference and statistical mechanics, from which the above transform is derived. If one equates the parameters to microstates $i$, the negative log-likelihood to the microstate energy $E_i$, and the prior to the density of states $g_i$, then the posterior as given by the generalised Bayes' rule is the canonical ensemble
\begin{equation}
    p(E_i) = \frac{g_i e^{-\beta E_i}}{Z(\beta)} \quad \leftrightarrow \quad \mathcal{P}_\beta(\theta) = \frac{\mathcal{L}^{\beta}(\theta)\pi(\theta)}{\mathcal{Z}(\beta)}
\end{equation}
at the inverse temperature $\beta = 1/T$. As noted by \citet{demons}, thermal algorithms such as thermodynamic integration \citep{path_sampling} get the evidence by evolving through a series of canonical ensembles via some temperature schedule, but nested sampling instead maintains \textit{microcanonical} ensembles, which are the iso-likelihood contours. Instead of using temperature \citep{simulated_annealing, parallel_tempering} or energy \citep{wang_landau} as a control parameter, nested sampling chooses a series of ensembles with constant relative volume entropy $\Delta \log X$, which allows the algorithm to handle phase transitions \citep{baldock}.
\par
Because the temperature of a microcanonical ensemble is a derived property rather than a parameter, there is some freedom in its definition. Returning to our original motivation, we make the connection that the temperature is the re-weighting $\Like \to \Like^{\beta}$ which centres the ensemble around the current energy. We now present several temperatures that achieve this aim, each of which one can plausibly consider to be the current temperature of a nested sampling run.

\subsubsection{Microcanonical temperature}\label{sec:microcanonical_temperature}
The obvious candidate is the microcanonical temperature $\partial S/\partial E$, where the volume entropy is $\log X$ and the energy is as usual $-\log \Like$. This gives the density of states; as discussed in Skilling's original paper,
\begin{equation}
    \beta_\mathrm{M}  = - \frac{\mathrm{d} \log X}{\mathrm{d} \log \Like} \Bigg\vert_{\log \Like_i}.
\end{equation}
is the $\beta$ at which $\Like^{\beta} X$ peaks at $\log X_i$, if we assume differentiability, which is exactly the intuition we were aiming for to put the ensemble bulk at the current contour. 
\par
Its value can be easily obtained via finite difference of the $\log \Like$ and $\log X$ intervals, albeit subject to an arbitrary window size for the differencing. Indeed, material science applications \citep{Baldock_2017} use this estimator to monitor the `cooling' progress of nested sampling, with a window size of 1000 iterations.

\subsubsection{Canonical temperature}\label{sec:canonical_temperature}
Another temperature considered by \citet{demons} is that at which the current energy (i.e. $-\log \Like_i$) is the average energy of the entire ensemble. One can obtain it by inverting 
\begin{equation}
    \langle \log \Like \rangle_{\mathcal{P}_\beta} = \log \Like_i
\end{equation}
to get the `canonical' temperature $\beta_\mathrm{C}$. While $\beta_\mathrm{M}$ is derived from (the gradient of) a single contour, this temperature uses the entire ensemble. It has the desirable property that it rises monotonically with compression, in analogy to a monotonic annealing schedule.

\subsubsection{Bayesian temperature}\label{sec:bayesian_temperature}
We furthermore propose a temperature $\beta_\mathrm{B}$ that is obtained via Bayesian inference, which returns a distribution rather than a point estimate. Since each value of $\beta$ leads to a different likelihood $\Like^{\beta}$, one can consider the posterior distribution as a function of $\log X$ to be \textit{conditioned} on $\beta$. We can therefore write
\begin{equation}
   \mathcal{P}(\log X \mid \beta) = \frac{\Like^{\beta}(X) X}{\mathcal{Z(\beta)}}.
\end{equation}
What we would really like is the distribution of $\beta$ at the present iteration, so the natural step is to invert this via Bayes' rule;
\begin{equation}
    P\left(\beta \mid \log X_i\right) = \frac{\mathcal{P}\left(\log X_i \mid \beta\right) P\left(\beta\right)}{P\left(\log X\right)}.
\end{equation}
As with all Bayesian analyses, the distribution of $\beta$ is fixed up to a prior, which we choose to be uniform in $\beta$. The obtained temperatures are consistent with the previous two choices, which may seem oddly coincidental. In fact, closer inspection reveals that large values of $P(\beta \mid \log X_i)$ are the temperatures with a large value of the posterior at the present contour, normalised by the corresponding evidence. Thus the Bayesian temperature uses the same idea as the microcanonical one, except it accounts for the spread in the result.

\subsubsection*{Comparisons}
\begin{figure}
\begin{center}
    \includegraphics{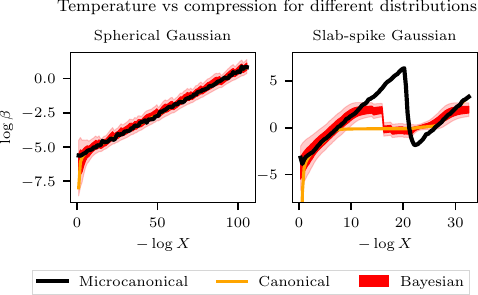}
\end{center}
\caption{Inferred temperatures using the microcanonical, canonical and Bayesian definitions. The shaded regions show the $1-2\sigma$ uncertainties. All are consistent for a single phase, but differ during a phase transition.}
\label{fig:beta_comparison}
\end{figure}

\cref{fig:beta_comparison} shows the three temperatures as a function of compression for two cases, one containing a phase transition and one without. They are consistent in both cases when there is a single dominant phase, but differ during a phase transition. The canonical temperature is the only one that rises monotonically with compression.
\par
One should keep in mind that despite the above theoretical reasoning, our introduction of the likelihood transformation was ultimately motivated by our wish to utilise the extra degree of freedom it provides. As we will see below, we recommend choosing the exact definition depending on what is useful for the problem at hand.

\subsection{Dimensionality}\label{sec:dimensionality}
We can immediately use the inferred temperature to track how the effective dimensionality of the posterior changes throughout the run, which was previously inaccessible. \citet{Handley_2019} demonstrated that at the end of a run, a measure of the number of constrained parameters is given by the Bayesian model dimensionality (BMD), defined as the posterior variance of the information content:
\begin{equation}
\frac{d_G}{2} = \int \mathcal{P}(\theta) \left(\log \frac{\mathcal{P}(\theta)}{\pi(\theta)} - \DKL\right)^2 \: \mathrm{d}\theta
= \langle \mathcal{I}^2 \rangle_\mathcal{P} - \langle \mathcal{I} \rangle^2_\mathcal{P}.
\end{equation}
\par
\begin{figure}
\begin{center}
    \includegraphics{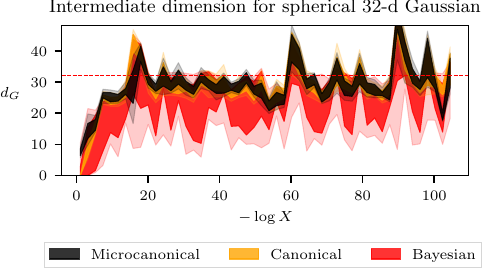}
\end{center}
\caption{Dimensionality estimates using the different temperatures for a spherical 32-d Gaussian. Again, all are consistent, but the Bayesian definition has a uncertainty which includes the true value far more consistently than for the other definitions.}
\label{fig:d_G_spherical}
\end{figure}

Calculating the quantity using intermediate set of weighted samples (which is concentrated at a single point) leads to vanishing variance, hence also dimensionality. However, we can recover the structure of the posterior together with the true dimensionality by adjusting the temperature. Dimensionality estimates are plotted in \cref{fig:d_G_spherical} for a spherical 32-d Gaussian, for which the true dimensionality is known. 
\par
The different choices of temperature are again consistent, but for the rest of this paper we choose the Bayesian $\beta$, because it provides a better reflection of the uncertainty in the estimate; the others, while fluctuating around the true value, are often many standard errors away from the true value at each single point.

\subsubsection*{Anisotropic compression}\label{sec:elongated}
\begin{figure}
\begin{center}
    \includegraphics{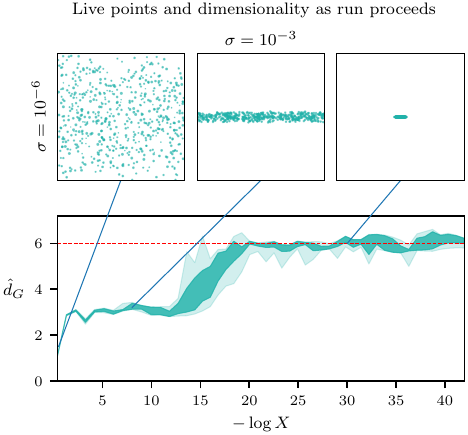}
\end{center}
\caption{Dimensionality estimates for a Gaussian that is elongated in half of its dimensions, with $1-2\sigma$ uncertainties shaded. The locations of the live points in the prior are shown at three stages, indicated by the connecting lines. As can be seen from the live point distribution, the prior does not compress in the higher variance direction until much later in the run, and early on it appears as if those directions are completely unconstrained.}
\label{fig:d_G_elongated}
\end{figure}

Plots of samples dimensionality against compression also draw attention to the \textit{directions} in which the samples are constrained throughout the run. As a concrete example, consider an elongated Gaussian in a unit hypercube prior with $\mu = \bm{0}$ and $\Sigma = \mathrm{diag}\left(10^{-3}, 10^{-3}, 10^{-3}, 10^{-6}, 10^{-6}, 10^{-6}\right)$, for which the dimensionality estimates are plotted in \cref{fig:d_G_elongated}. Alongside is a view of the distribution of live points across the prior for two directions with different scales, which shows the level to which those parameters have been constrained at different times.
\par
A feature of nested sampling made apparent here is that parameters with high variance are initially `hidden'. Compression occurs in the direction which is most likely to have a sample of higher likelihood, and initially it is much easier to find a better point along the direction of a parameter that is poorly constrained. Lower variance parameters are constrained much later, and before that happens it appears as though those parameters have a uniform distribution.
\par
It is important to appreciate that at lower compression the samples truly lie in a lower-dimensional space, rather than some artefact of the way we view them. Anticipating the full dimensionality of the space is therefore just as impossible as that associated with a slab-spike geometry, so in this sense such geometries contain a `compressive phase transition'. 

\section{Endpoint prediction}\label{sec:endpoint}
As described in \citet{supernest} and further explored in the talk and upcoming paper \citet{kcl_talk, scaling_frontier}, the time complexity of nested sampling is
\begin{equation}
    T \propto \langle \tfrac{1}{\nlive} \rangle^{-1} \times \langle \mathcal{T}\{ \Like(\theta) \} \rangle \times \langle f_\mathrm{sampler} \rangle \times \DKL.
\end{equation}
The first term is the harmonic mean of the number of live points $\sim\mathcal{O}(n_i)$, The second term is the average time per likelihood evaluation. The third is the average number of evaluations required to replace a dead point with a live point at higher likelihood, which is given by the implementation and usually does not vary in orders of magnitude. 
The final term is the Kullback-Liebler divergence~\cref{eq:DKL}, the compression factor required to get from the prior to the posterior. This term is generally outside of user control, in most cases \textit{a priori} unknown, and of principle interest in this section.

\subsection{The termination prior volume}
Making the above discussion more precise, we wish to find the compression factor $\log X_ \mathrm{f}$ at which the termination criterion is met, which is larger in magnitude than $\DKL$ (\cref{fig:logX_distribution}). The difficulty is that at an intermediate iteration we only know the posterior up to the maximum log-likelihood live point, which until just before the end is far from the posterior bulk. 
\par
In order to get an idea of where the true posterior bulk sits, we need to predict what the posterior looks like past the highest live point. We do this by \textit{extrapolating} the known likelihood profile; that is, the trajectory of $\Like(X)$ traced out by the live and dead points. 
One would never use this predicted posterior to perform inference, since more accuracy can always be achieved by simply finishing the run. However, we will demonstrate it is sufficient for making a run-time prediction for $\log X_\mathrm{f}$. 
\par
Quantitatively, this proceeds as follows: fit a function $f(X, \phi)$  with some parameters $\phi$ to the known likelihood profile, which allows us to express the prior volume we need to compress to as
\begin{equation}
	\Delta \mathcal{Z} = \epsilon \mathcal{Z}_\mathrm{tot},
\end{equation}
or equivalently
\begin{equation}\label{endpoint}
	\int_0^{X_\mathrm{f}} f(X, \phi)\ \mathrm{d}X = \epsilon \left( \int_0^{X_i} f(X, \phi)\ \mathrm{d}X + \mathcal{Z}_\mathrm{dead} \right),
\end{equation}
where $X_i$ is the volume of the iteration we have currently compressed to, and $\mathcal{Z}_\mathrm{dead}$ is the evidence we have accumulated up to this point. $X_\mathrm{f}$ can then be identified by solving the above equation either analytically or numerically. 
\par
Once $X_\mathrm{f}$ is known, the corresponding iteration count depends on the live point schedule. For example, in the constant $\nlive$ case $\log X$ decreases by $1/n$ at each iteration, so the total number of iterations $N_\mathrm{f}$ would be
\begin{equation}
	N_\mathrm{f} = - n\log X_\mathrm{f} .
\end{equation}

\subsection{How to extrapolate?}
A key observation is that the Bayesian model dimensionality is the equivalent dimension of the posterior if it were actually Gaussian. Fitting a Gaussian of this dimension to the likelihood profile therefore makes a reasonable approximation to the true distribution, without explicitly assuming the form of the likelihood function. The parameterisation of the Gaussian that we fit is the same as that given in \cref{sec:logL}, which we shall repeat here for clarity;
\begin{equation}\label{gaussian}
    f(X; \phi) = \logLmax - X^{2/d}/2\sigma^2
\end{equation}
The extrapolation then proceeds thus:
\begin{enumerate}[leftmargin=*]
    \item Find the current dimensionality ${d}^{(i)}_G$ of the posterior at the Bayesian temperature
    \item Take the live point profile and perform a least squares fit to \eqref{gaussian}, stipulating that $d = {d}^{(i)}_G$ to infer $\logLmax$ and $\sigma$ 
    \item Use the likelihood predicted by these parameters to solve \eqref{endpoint} for $X_\mathrm{f}$
\end{enumerate}
The advantage of fitting a Gaussian is that the procedure can be sped up analytically. Firstly, the least squares regression is trivial because analytic estimators exist; the cost function 
\begin{equation}\label{chi squared}
	C^2(\logLmax, \sigma) = \sum_i \left| \log \Like_i - f(X_i; \logLmax, \sigma) \right| ^2
\end{equation}
is minimised with respect to $(\logLmax, \sigma)$ when
\begin{equation}\label{eq:sigma}
    \sigma^2 = \frac{N \sum_i X_i^{4/d} - \left(\sum_i X_i^{2/d}\right)^2}{2 \sum_i \log \Like_i \sum_i X_i^{2/d} - 2N \sum_i X_i^{2/d}\log \Like_i },
\end{equation}
and
\begin{equation}\label{eq:logLmax}
    \logLmax = \frac{1}{N} \sum_i \log \mathcal{L}_i + \frac{1}{2N\sigma^2} \sum_i X_i^{2/d}.
\end{equation}
Secondly, the termination prior volume can also be obtained analytically. Rewriting \cref{endpoint} in terms of the Gaussian parameters gives
\begin{equation}
	\epsilon = \frac{\int_0^{X_\mathrm{f}} \Like_\mathrm{max} \exp\left(-X^{2/d}/2\sigma^2\right)\ \mathrm{d}X}{\int_0^{X_i} \Like_\mathrm{max} \exp\left(-X^{2/d}/2\sigma^2\right)\ \mathrm{d}X + \mathcal{Z}_\mathrm{dead}}.
\end{equation}
The integrals have the analytic solution
\begin{equation}
	\int_0^{X_k} \Like_\mathrm{max} \exp\left(-X^{2/d}/2\sigma^2\right)\ \mathrm{d}X = \frac{d}{2} \cdot \left(\sqrt{2}\sigma\right)^d \cdot \gamma_k
\end{equation}
where $\gamma_k = \Gamma_{d/2}\left(X_k^{2/d}/2\sigma^2\right)$ is the lower incomplete gamma function. After taking the inverse of  $\gamma$ and a few more steps of algebra, we arrive at
\begin{equation}
    \log X_\mathrm{f} = \frac{d}{2}\log 2\sigma^2	+ \log \Gamma^{-1}_{d/2} \left(\epsilon \gamma_i+ \frac{\epsilon\mathcal{Z}_\mathrm{dead}}{ \left( 2\sigma^2 \right)^{d/2}\Like_\mathrm{max}}\right),\label{eq:xf}
\end{equation}
and $N_\mathrm{f}$ is of course just $-n$ multiplied by this. Intuitively, the above procedure can be thought of as inferring the number of constrained parameters, then extrapolating them up to find the point at which they will be fully constrained. 
\par
Uncertainties in the final estimate are obtained by drawing many samples from the distribution of $d_\mathrm{G}$ defined by the Bayesian temperature, and repeating step two for each. One might wonder why we do not obtain $d$ via least squares regression together with the other parameters; extensive testing has shown this approach to be far less stable. 

\subsection{Alternative approaches}
More comprehensive Bayesian approaches, perhaps including \textit{a priori} information about the likelihood or greater flexibility in the fitting function, could likely perform better than what we have just presented. However, such methods would not befit \textit{run-time} prediction which has a much more limited computational budget, hence the more pragmatic approach we have adopted. Here, we discuss as a benchmark alternative approaches to endpoint estimation that have a comparable computational complexity. 
\subsubsection{Integral progress}
An alternative approach used in \textsc{Ultranest} \citep{ultranest} derives a progress bar based on the fraction of the accumulated integral compared to the remaining integral, approximated as 
\begin{equation}\label{eq:integral_progress}
    \mathcal{Z}_\mathrm{rem} \approx \Like_\mathrm{max}^{\mathrm{live}} X_i.
\end{equation}
This has a several shortcomings. First, run-time is proportional to compression rather than accumulation of the integral, since it takes just as long to traverse the width of the bulk as it does any other width. Second, because of the point-like nature of the posterior mid-run, the remaining integral approximated as such holds nearly all of the evidence, so the relative fraction of the accumulated and remaining evidence is almost zero for most of the run. Finally, approximation \eqref{eq:integral_progress} is always an underestimate, because as previously found the maximum live point is generally nowhere near the true maximum. This approach can however be useful in the low dimensions appropriate for \textsc{Ultranest} when the live points are always near the maximum, but in general is less reliable.

\subsubsection{Extrapolating evidence increments}
\begin{figure}
\begin{adjustbox}{valign=t, scale=0.72}
\subfloat{\begin{tabular}{|c|c|c|}
\hline
iteration & log Z  & $\Delta\log Z$   \\
\hline
5000 & -1435.8 & 190.8 \\
5500 & -1264.6 & 171.2 \\
6000 & -1123.7 & 140.9 \\
6500 & -991.5 & 132.2 \\
7000 & -885.0 & 106.6 \\
7500 & -790.3 & 94.7 \\
8000 & -702.6 & 87.7 \\
8500 & -619.7 & 82.9 \\
9000 & -551.8 & 67.9 \\
9500 & -492.7 & 59.1 \\
\hline
\end{tabular}}
\end{adjustbox}
\quad
\begin{adjustbox}{valign=t}
\subfloat{\includegraphics{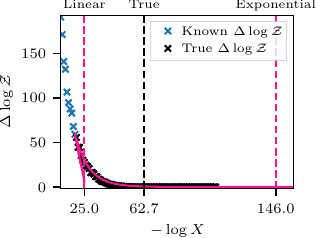}}
\end{adjustbox}
\caption{Extrapolating the increments of evidence. The left column shows the output of a nested sampling run, and the right column shows the extrapolation.}
\label{fig:inc_extrapolate}
\end{figure}

Seasoned watchers of nested sampling runs might be curious how the method compares to simply extrapolating the increments of evidence to roughly estimate when the evidence converges. We do this for a spherical Gaussian and compare it to our method. At an intermediate stage of the run, the most recent outputs might look something like that shown in the first two columns of the table in \cref{fig:inc_extrapolate}. Extrapolating those data to a linear and exponential profile yields endpoint estimates plotted in the graph to the right. 
\par
\begin{figure}
\includegraphics{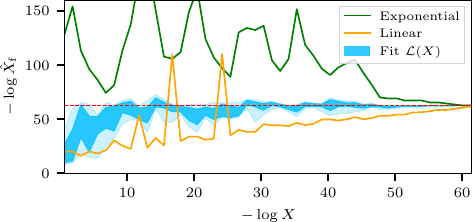}
\caption{Endpoint predictions for a spherical Gaussian, with $1-2\sigma$ uncertainties shaded. Extrapolating the evidence increments in a linear/exponential manner under/over-predicts the endpoint, both of which perform considerably worse than the method of extrapolating the likelihood.}
\label{fig:inc_predictions}
\end{figure}

The linear extrapolation is clearly an underestimate, since it fails to account for the long tail of the nonlinear profile. The increments are also not exactly exponential, since the exponential fit leads to a large over-prediction. The predicted endpoint over the course of a run for $d = 16$, $\sigma = 0.01$, as shown in \cref{fig:inc_predictions}, shows the same result. One might expect an average to be more accurate, but this tends to be biased towards the exponential prediction, and there is no obvious choice of weighting that would fix this.
\par
More importantly, we find that for real likelihoods which have an element of noise the extrapolation often diverges, for instance when the increments do not monotonically decrease. Directly extrapolating the evidence increments is therefore far less stable than the previous method, and generally not a reliable method for prediction.

\section{Results}\label{sec:results}
We now test the approach established in the previous section on a range of distributions. We begin by considering a series of toy examples to explore the capabilities and limitations of the method, before presenting results for real cosmological chains.
\subsection{Toy examples}
Throughout the toy examples we make use the perfect nested sampling~\citep{Keeton_2011,2018BayAn..13..873H} framework implemented in \textsc{anesthetic}~\citep{anesthetic}.
\subsubsection{Gaussians}
\begin{figure*}
\begin{center}
    \includegraphics{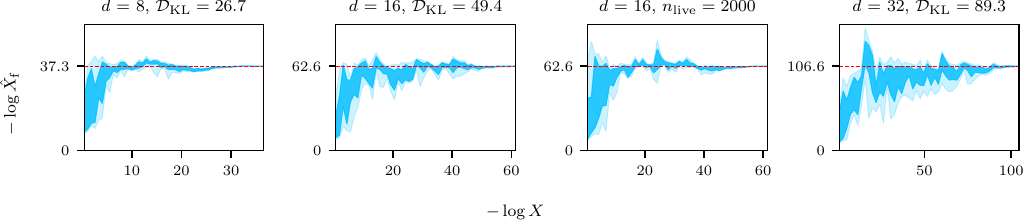}
\end{center}
\caption{Endpoint predictions for a spherical Gaussian run with $\nlive=500$ (except for the third plot from left). The correct endpoint is obtained for all but the earliest iterations, and the uncertainty is controlled by the number of live points, which can be seen from the two $d = 16$ plots.}
\label{fig:gauss_predictions}
\end{figure*}

Predictions for spherical Gaussians of various dimensions are shown in \cref{fig:gauss_predictions} as a benchmark for when fitting a Gaussian distribution is exact. In all endpoint prediction plots, the shading indicates the $1-2\sigma$ uncertainties. All were run with $\nlive = 500$ except for one $\nlive=2000$ for comparison, with each Gaussian having a width of $\sigma = 0.01$. The correct endpoint is recovered to within standard error at all points except the very beginning, when the parameters have hardly been constrained. 
\par
\begin{figure}
\begin{center}
    \includegraphics{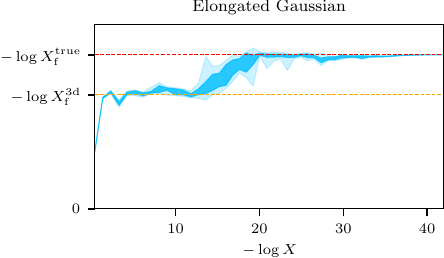}
\end{center}
\caption{Endpoint prediction for an elongated Gaussian. At early stages, the full dimensionality is undetectable, and the endpoint is predicted to be the same as for a likelihood with three unconstrained directions. Only once the prior has been compressed enough to constrain the other three directions does the prediction converge to the true value.}
\label{fig:elongated_logXfs}
\end{figure}

We note that as with nested sampling in general, increasing $\nlive$ improves the resolution and reliability of the inferences, which can be seen from the middle two plots.
We also observe the effect of elongating the Gaussian, using the same example as \cref{sec:elongated}. \cref{fig:elongated_logXfs} shows a step-like trend similar to the inferred dimensionalities, reflecting the fact that the full dimensionality is undetectable at lower compression factors. The endpoint for a likelihood whose remaining three directions are completely unconstrained coincides with our predictions at early iterations, showing that the two cases are indistinguishable. 

\subsubsection{Cauchy}
\begin{figure*}
\begin{center}
    \includegraphics{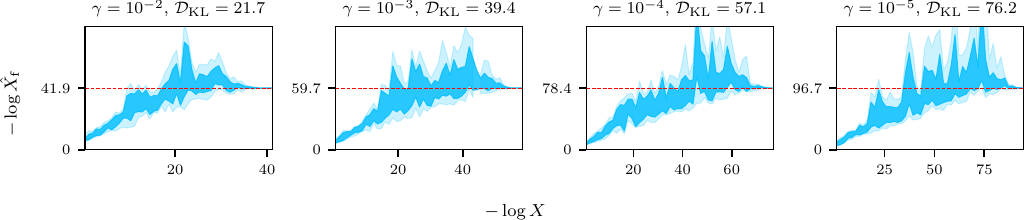}
\end{center}
\caption{Predictions for the Cauchy distribution for various widths $\gamma$ and $\nlive = 500$. The endpoint is underestimated for the first half of the run, but this is a limitation of the Gaussian approximation rather than a lack of information mid-run.}

\label{fig:cauchy_predictions}
\end{figure*}

One case that might be expected to cause problems is the pathological Cauchy distribution, which is far from a Gaussian. \cref{fig:cauchy_predictions} shows the predictions for a likelihood of the form
\begin{equation}
	\log\mathcal{L} = \log\mathcal{L}_\mathrm{max} - \frac{1 + d}{2} \log \left(1 + \frac{X^2}{\gamma^2}\right),
\end{equation}
choosing $d = 10$ and allowing $\gamma$ to vary. The correct estimate is obtained to within standard error by about halfway, but before that is inaccurate. The key limitation is that the estimate is wrong early on, not because the compression is anisotropic, or because there is a phase transition; but rather as a limitation of the reducing the likelihood to a Gaussian via the BMD, which is itself less stable for a Cauchy.
\par
Nevertheless, the right order of magnitude is obtained at all times, so this remains sufficient for most use-cases. The Cauchy is also a pathological case, and the same problem does not in practice appear for more realistic cases, as we shall see next.

\subsection{Cosmological examples}
\begin{figure*}
\begin{center}
    \includegraphics{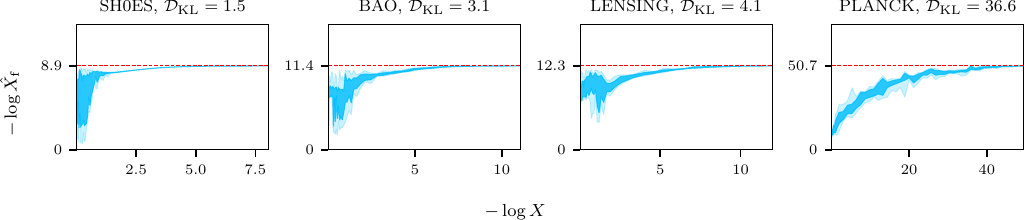}
\end{center}
\caption{Endpoint predictions for cosmological likelihoods. The first three low $\DKL$ inferences get the correct endpoint from the start, while the \textsc{Planck} chain takes longer to converge because the likelihood is highly covariant and thus subject to anisotropic compression, which makes the samples lie in a lower dimensional subspace mid-run.}
\label{fig:lcdm_logXfs}
\end{figure*}

Finally, we evaluate the method on real cosmological chains. \cref{fig:lcdm_logXfs} presents the endpoints (calculated after the fact) for nested sampling runs for curvature quantification on several common cosmological data sets (details in \citealt{curvature_tension}). 
\par
The \textsc{SH0ES}, \textsc{BAO} and \textsc{lensing} chains are `easy' low $\DKL$ inferences, so it is expected that the correct endpoint is inferred practically from the start. The \textsc{Planck} endpoints, on the other hand, are not correct until at least midway through. However, this is expected from the covariance of the \textsc{Planck} likelihood, which consists of principal components of many scales and therefore elongated in many dimensions. It is therefore of the same class as the elongated Gaussian presented in \cref{sec:elongated}; the samples exist in a lower dimensional subspace mid-run, which slowly increases to the full dimensionality only at the end of the run.

\section{Conclusion}
We have derived new analytic results to make an anatomy of nested sampling, understanding the progression of a run via the compression of prior volume, the increase in log-likelihood, the inferred temperature schedule, and the convergence of the sample dimensionality. From these analyses, we developed a method for predicting the endpoint of a nested sampling run, by using the inferred Bayesian model dimensionality mid-run to extrapolate the known likelihood profile. 
\par
The method in general converges on a correct prediction of endpoint by about halfway, and gets the correct order of magnitude throughout. Consistent predictions are obtained for both toy and cosmological examples. The accuracy is typically limited by the information available mid-run, either because of a phase transition or because the anisotropy of the nested sampling compression. Pathological distributions, such as a Cauchy, lead to less stable inferences of the dimensionality and expose the limitations of a Gaussian approximation, though the order of magnitude is still correct.
\par
Further work can be done to experiment with more flexible basis functions for regression of the likelihood profile, so that it is less dependent on the Gaussian approximation.

\section*{Code availability}
A package is in development to implement the endpoint prediction mechanism, for easy plug-in to existing implementations of nested sampling. The latest updates can be found at \href{https://github.com/zixiao-h/aeons}{github.com/zixiao-h/aeons}.

\section*{Acknowledgements}
WJH is supported by the Royal Society as a Royal Society University Research Fellow at the University of Cambridge. AB was supported by a Cambridge Mathematical Placement and the Royal Society summer studentship. ZH was supported by a Royal Society summer studentship. We thank Mike Hobson for helpful discussion during ZH Part III viva.

\bibliographystyle{mnras}
\bibliography{references}

\begin{thebibliography}{}
\makeatletter
\relax
\def\mn@urlcharsother{\let\do\@makeother \do\$\do\&\do\#\do\^\do\_\do\%\do\~}
\def\mn@doi{\begingroup\mn@urlcharsother \@ifnextchar [ {\mn@doi@}
  {\mn@doi@[]}}
\def\mn@doi@[#1]#2{\def\@tempa{#1}\ifx\@tempa\@empty \href
  {http://dx.doi.org/#2} {doi:#2}\else \href {http://dx.doi.org/#2} {#1}\fi
  \endgroup}
\def\mn@eprint#1#2{\mn@eprint@#1:#2::\@nil}
\def\mn@eprint@arXiv#1{\href {http://arxiv.org/abs/#1} {{\tt arXiv:#1}}}
\def\mn@eprint@dblp#1{\href {http://dblp.uni-trier.de/rec/bibtex/#1.xml}
  {dblp:#1}}
\def\mn@eprint@#1:#2:#3:#4\@nil{\def\@tempa {#1}\def\@tempb {#2}\def\@tempc
  {#3}\ifx \@tempc \@empty \let \@tempc \@tempb \let \@tempb \@tempa \fi \ifx
  \@tempb \@empty \def\@tempb {arXiv}\fi \@ifundefined
  {mn@eprint@\@tempb}{\@tempb:\@tempc}{\expandafter \expandafter \csname
  mn@eprint@\@tempb\endcsname \expandafter{\@tempc}}}

\bibitem[\protect\citeauthoryear{Akrami, Scott, Edsjö, Conrad  \&
  Bergström}{Akrami et~al.}{2010}]{Akrami_2010}
Akrami Y.,  Scott P.,  Edsjö J.,  Conrad J.,   Bergström L.,  2010, \mn@doi
  [Journal of High Energy Physics] {10.1007/jhep04(2010)057}, 2010

\bibitem[\protect\citeauthoryear{Ashton et~al.,}{Ashton
  et~al.}{2022}]{physical_scientists}
Ashton G.,  et~al., 2022, \mn@doi [Nature Reviews Methods Primers]
  {10.1038/s43586-022-00121-x}, 2

\bibitem[\protect\citeauthoryear{Baldock, P\'artay, Bart\'ok, Payne  \&
  Cs\'anyi}{Baldock et~al.}{2016}]{baldock}
Baldock R. J.~N.,  P\'artay L.~B.,  Bart\'ok A.~P.,  Payne M.~C.,   Cs\'anyi
  G.,  2016, \mn@doi [Phys. Rev. B] {10.1103/PhysRevB.93.174108}, 93, 174108

\bibitem[\protect\citeauthoryear{Baldock, Bernstein, Salerno, P{\'{a}}rtay  \&
  Cs{\'{a}}nyi}{Baldock et~al.}{2017}]{Baldock_2017}
Baldock R. J.~N.,  Bernstein N.,  Salerno K.~M.,  P{\'{a}}rtay L.~B.,
  Cs{\'{a}}nyi G.,  2017, \mn@doi [Physical Review E]
  {10.1103/physreve.96.043311}, 96

\bibitem[\protect\citeauthoryear{{Brewer} \& {Foreman-Mackey}}{{Brewer} \&
  {Foreman-Mackey}}{2016}]{dnest}
{Brewer} B.~J.,  {Foreman-Mackey} D.,  2016, \mn@doi [arXiv e-prints]
  {10.48550/arXiv.1606.03757}, \href
  {https://ui.adsabs.harvard.edu/abs/2016arXiv160603757B} {p. arXiv:1606.03757}

\bibitem[\protect\citeauthoryear{Buchner}{Buchner}{2021}]{ultranest}
Buchner J.,  2021, UltraNest -- a robust, general purpose Bayesian inference
  engine (\mn@eprint {arXiv} {2101.09604})

\bibitem[\protect\citeauthoryear{Buchner}{Buchner}{2023}]{Buchner_2023}
Buchner J.,  2023, \mn@doi [Statistics Surveys] {10.1214/23-ss144}, 17

\bibitem[\protect\citeauthoryear{Chopin \& Robert}{Chopin \&
  Robert}{2010}]{Chopin_2010}
Chopin N.,  Robert C.~P.,  2010, \mn@doi [Biometrika] {10.1093/biomet/asq021},
  97, 741

\bibitem[\protect\citeauthoryear{Evans}{Evans}{2006}]{evans}
Evans M.,  2006, World Meeting on Bayesian Statistics Benidorm, 8

\bibitem[\protect\citeauthoryear{Feroz, Hobson  \& Bridges}{Feroz
  et~al.}{2009}]{multinest}
Feroz F.,  Hobson M.~P.,   Bridges M.,  2009, \mn@doi [Monthly Notices of the
  Royal Astronomical Society] {10.1111/j.1365-2966.2009.14548.x}, 398, 1601

\bibitem[\protect\citeauthoryear{Feroz, Cranmer, Hobson, de Austri  \&
  Trotta}{Feroz et~al.}{2011}]{Feroz_2011}
Feroz F.,  Cranmer K.,  Hobson M.,  de Austri R.~R.,   Trotta R.,  2011,
  \mn@doi [Journal of High Energy Physics] {10.1007/jhep06(2011)042}, 2011

\bibitem[\protect\citeauthoryear{Gelman \& Meng}{Gelman \&
  Meng}{1998}]{path_sampling}
Gelman A.,  Meng X.-L.,  1998, \mn@doi [Statistical Science]
  {10.1214/ss/1028905934}, 13, 163

\bibitem[\protect\citeauthoryear{Habeck}{Habeck}{2015}]{demons}
Habeck M.,  2015, \mn@doi [AIP Conference Proceedings] {10.1063/1.4905971},
  1641, 121

\bibitem[\protect\citeauthoryear{Handley}{Handley}{2019}]{anesthetic}
Handley W.,  2019, \mn@doi [Journal of Open Source Software]
  {10.21105/joss.01414}, 4, 1414

\bibitem[\protect\citeauthoryear{Handley}{Handley}{2021}]{curvature_tension}
Handley W.,  2021, \mn@doi [Physical Review D] {10.1103/physrevd.103.l041301},
  103

\bibitem[\protect\citeauthoryear{Handley}{Handley}{2023a}]{kcl_talk}
Handley W.,  2023a, Nested sampling: powering next-generation inference and
  machine learning tools for cosmology, particle physics and beyond, King's
  College London, \url
  {https://github.com/williamjameshandley/talks/raw/kcl_2023/will_handley_kcl_2023.pdf}

\bibitem[\protect\citeauthoryear{Handley}{Handley}{2023b}]{scaling_frontier}
Handley W.,  2023b, The scaling frontier of nested sampling

\bibitem[\protect\citeauthoryear{Handley \& Lemos}{Handley \&
  Lemos}{2019}]{Handley_2019}
Handley W.,  Lemos P.,  2019, \mn@doi [Physical Review D]
  {10.1103/physrevd.100.023512}, 100

\bibitem[\protect\citeauthoryear{Handley, Hobson  \& Lasenby}{Handley
  et~al.}{2015}]{polychord}
Handley W.~J.,  Hobson M.~P.,   Lasenby A.~N.,  2015, \mn@doi [Monthly Notices
  of the Royal Astronomical Society] {10.1093/mnras/stv1911}, 453, 4385

\bibitem[\protect\citeauthoryear{Higson, Handley, Hobson  \& Lasenby}{Higson
  et~al.}{2018a}]{sparse_reconstruction}
Higson E.,  Handley W.,  Hobson M.,   Lasenby A.,  2018a, \mn@doi [Monthly
  Notices of the Royal Astronomical Society] {10.1093/mnras/sty3307}

\bibitem[\protect\citeauthoryear{{Higson}, {Handley}, {Hobson}  \&
  {Lasenby}}{{Higson} et~al.}{2018b}]{2018BayAn..13..873H}
{Higson} E.,  {Handley} W.,  {Hobson} M.,   {Lasenby} A.,  2018b, \mn@doi
  [Bayesian Analysis] {10.1214/17-BA1075}, \href
  {https://ui.adsabs.harvard.edu/abs/2018BayAn..13..873H} {13, 873}

\bibitem[\protect\citeauthoryear{Higson, Handley, Hobson  \& Lasenby}{Higson
  et~al.}{2019}]{dynamic_ns}
Higson E.,  Handley W.,  Hobson M.,   Lasenby A.,  2019, \mn@doi [Statistics
  and Computing] {10.1007/s11222-018-9844-0}, 29, 891–913

\bibitem[\protect\citeauthoryear{Keeton}{Keeton}{2011}]{Keeton_2011}
Keeton C.~R.,  2011, \mn@doi [Monthly Notices of the Royal Astronomical
  Society] {10.1111/j.1365-2966.2011.18474.x}, 414, 1418

\bibitem[\protect\citeauthoryear{Kirkpatrick, Gelatt  \& Vecchi}{Kirkpatrick
  et~al.}{1983}]{simulated_annealing}
Kirkpatrick S.,  Gelatt C.~D.,   Vecchi M.~P.,  1983, \mn@doi [Science]
  {10.1126/science.220.4598.671}, 220, 671

\bibitem[\protect\citeauthoryear{{McEwen}, {Liaudat}, {Price}, {Cai}  \&
  {Pereyra}}{{McEwen} et~al.}{2023}]{proxnest}
{McEwen} J.~D.,  {Liaudat} T.~I.,  {Price} M.~A.,  {Cai} X.,   {Pereyra} M.,
  2023, \mn@doi [arXiv e-prints] {10.48550/arXiv.2307.00056}, \href
  {https://ui.adsabs.harvard.edu/abs/2023arXiv230700056M} {p. arXiv:2307.00056}

\bibitem[\protect\citeauthoryear{Petrosyan \& Handley}{Petrosyan \&
  Handley}{2022}]{supernest}
Petrosyan A.,  Handley W.,  2022, \mn@doi [Physical Sciences Forum]
  {10.3390/psf2022005051}, 5

\bibitem[\protect\citeauthoryear{Pártay, Bartók  \& Csányi}{Pártay
  et~al.}{2010}]{statmech}
Pártay L.~B.,  Bartók A.~P.,   Csányi G.,  2010, Efficient sampling of
  atomic configurational spaces (\mn@eprint {arXiv} {0906.3544})

\bibitem[\protect\citeauthoryear{Pártay., Csányi  \& Bernstein}{Pártay.
  et~al.}{2021}]{materials}
Pártay. L.~B.,  Csányi G.,   Bernstein N.,  2021, \mn@doi [The European
  Physical Journal B] {10.1140/epjb/s10051-021-00172-1}, 94

\bibitem[\protect\citeauthoryear{Sivia \& Skilling}{Sivia \&
  Skilling}{2006}]{sivia}
Sivia D.~S.,  Skilling J.,  2006, Data Analysis: A bayesian tutorial (Oxford
  Science Publications).
Oxford University Press

\bibitem[\protect\citeauthoryear{Skilling}{Skilling}{2006}]{skilling}
Skilling J.,  2006, \mn@doi [Bayesian Analysis] {10.1214/06-ba127}, 1

\bibitem[\protect\citeauthoryear{Speagle}{Speagle}{2020}]{dynesty}
Speagle J.~S.,  2020, \mn@doi [Monthly Notices of the Royal Astronomical
  Society] {10.1093/mnras/staa278}, 493, 3132

\bibitem[\protect\citeauthoryear{Swendsen \& Wang}{Swendsen \&
  Wang}{1986}]{parallel_tempering}
Swendsen R.~H.,  Wang J.-S.,  1986, \mn@doi [Phys. Rev. Lett.]
  {10.1103/PhysRevLett.57.2607}, 57, 2607

\bibitem[\protect\citeauthoryear{Trotta, Feroz, Hobson, Roszkowski  \& de
  Austri}{Trotta et~al.}{2008}]{Trotta_2008}
Trotta R.,  Feroz F.,  Hobson M.,  Roszkowski L.,   de Austri R.~R.,  2008,
  \mn@doi [Journal of High Energy Physics] {10.1088/1126-6708/2008/12/024},
  2008, 024

\bibitem[\protect\citeauthoryear{Wang \& Landau}{Wang \&
  Landau}{2001}]{wang_landau}
Wang F.,  Landau D.~P.,  2001, \mn@doi [Phys. Rev. Lett.]
  {10.1103/PhysRevLett.86.2050}, 86, 2050

\bibitem[\protect\citeauthoryear{{Williams}, {Veitch}  \&
  {Messenger}}{{Williams} et~al.}{2021}]{nessai}
{Williams} M.~J.,  {Veitch} J.,   {Messenger} C.,  2021, \mn@doi [Physical
  Review D] {10.1103/PhysRevD.103.103006}, \href
  {https://ui.adsabs.harvard.edu/abs/2021PhRvD.103j3006W} {103, 103006}

\makeatother
\end{thebibliography}

\label{lastpage}
\end{document}